# Enhanced solar evaporation of water from porous media, through capillary mediated forces and surface treatment

F. M. Canbazoglu, B. Fan, A. Kargar, K. Vemuri, and P.R. Bandaru[*]

Department of Mechanical Engineering, University of California, San Diego, La Jolla, CA

**Abstract**

The relative influence of the capillary, Marangoni, and hydrophobic forces in mediating the evaporation of water from carbon foam based porous media, in response to incident solar radiation, are investigated. It is indicated that inducing hydrophilic interactions on the surface, through nitric acid treatment of the foams, has a similar effect to reduced pore diameter and the ensuing capillary forces. The efficiency of water evaporation may be parameterized through the Capillary number (*Ca*), with a lower *Ca* being preferred. The proposed study is of much relevance to efficient solar energy utilization.

---

[*] Email: pbandaru@ucsd.edu

The enhancement of the efficiency of evaporation of liquid water, for water evaporation and steam generation, is of major technological as well as scientific interest, with applications ranging from water heaters to distillation and desalination[1]. The use of solar radiation for this purpose amounts to a better utilization of an abundant energy resource[2,3]. While evaporation from planar surfaces is typically due to the differences between (a) the saturation vapor pressure and the internal liquid pressure, as well as (b) the temperature difference between the surface and the ambient, the use of porous media and heat localization may provide an alternate path, as has recently been proposed[4]. Such a scheme involves coupling the absorbed heat with capillary flow in porous media, where the "wicking" of the liquid in the channels provides a driving force for the liquid flow and subsequent evaporation. Much related work has focused on the chemical aspects of the process[5], as related to or enhanced light capture, *e.g.,* through the use of plasmonic characteristics[6,7] of nano-particles[8,9,10]/-shells[6,11,12]. However, considering that the efficacy of nanoparticles in enhancing convective heat transfer is debated[13], it would also be pertinent to consider the underlying mechanical and surface forces that may dictate the evaporation efficiency. It is then the aim of this letter to probe such fundamental aspects, related to the role of thermal and fluid transport[14], coupled with materials characteristics.

The experiments: Fig. 1(a), related to probing such characteristics were carried out, using carbon foam (from Reynolds Inc.) based porous media, with varying pore sizes (characterized in terms of pores per inch: PPI), albeit at a fixed porosity of 97%. We report on the results of particular samples with 45 PPI (with an average pore diameter, $d \sim 546$ $\mu$m), 60 PPI ($d \sim 422$ $\mu$m), 80 PPI ($d \sim 288$ $\mu$m), and 100 PPI ($d \sim 253$ $\mu$m) with a typical distribution (see supplementary material[29] in Section S1) of the pore sizes: Fig. 1(b) and *inset*. While the temperatures along the height of the aerogel insulated container – ranging from below the foam to above the foam (sampling the water vapor temperature) were monitored through appropriately placed thermocouples (TC), the evaporative heat flux was parameterized in terms of a mass loss/change ($\Delta m$) of the water from the beaker, which was measured through a well-calibrated mass balance. All the possible heat losses due to convective and radiative effects were carefully considered and calibrated. It was apparent from the results: Fig. 2(a), that $\Delta m$ was approximately linear with waiting time ($t$) and more pronounced at smaller pore sizes/average pore diameter ($d$). The rate of the mass loss ($=d(\Delta m)/dt$): Fig. 2(b), obtained through the derivatives of the curves in Fig. 2(a), indicates that a steady state was reached at longer times, with a larger change for smaller $d$. The steady state may be related to equilibrium between evaporation and the reverse process of condensation, *e.g.,* due to vapor buildup in the space above the sample. We also note that a reduced pore size resulted in a faster commencement of evaporation. In the present reported set of experiments, it was not specifically aimed to enhance the evaporative efficiency: $\eta = (\Delta m L)/t P_s$, indicating the mass loss of the liquid water transforming into vapor over a given time ($t$), with $L$ as the latent heat of evaporation, and $P_s$ as the input solar radiation power. Instead, the objective was to probe a relatively simple materials platform with respect to the thermal and mechanical forces. However, we estimate an efficiency of the order of ~ 60% from the presented data, with a $P_s$ of ~ 1700 W/m$^2$, provided by a solar simulator.

Generally, in solar evaporation involving porous media, there are several variables that govern the efficiency via the throughput and conversion of the liquid water, to water vapor. These include, *e.g.,* (a) the constitution and characteristics of the materials in the porous medium and related chemistry[15], (b) the pore size and the distribution, (c) the relative contribution of conductive and convective heat transfer, *etc*. The net *outwards* mass flux ($\Delta m$) across the interface (considering the difference of the evaporation and the condensation) has generally been considered to be a function of several intensive variables[16], in addition to the specific characteristics of the liquid (*e.g.,* the density: $\rho_l$, thermal conductivity: $\kappa_l$, dynamic viscosity: $\mu$, the surface tension: $\sigma$, the latent heat of evaporation: $L$), at/close to the interface such as the equilibrium vapor pressure (/temperature) at the liquid (*l*)/vapor (*v*) interface: $P_{lv}$ (/$T_{lv}$). At the interface of the liquid with the wall, the evaporation process was predominantly posited[16] to occur in an intermediate *evaporating thin film region* - between the *intrinsic meniscus* (*i.e.,* corresponding to the bulk) and a *non-evaporating* film - also see Fig. 4(b). The latter region has a thickness ($\delta$) of the order of ~ 10 nm, originates from the van der Waals forces between the solid surface and the adherent liquid, and creates a disjoining pressure: $P_d$ (~ $A/\delta^3$, with $A$ as the dispersion/Hamaker constant[17]). The $P_d$ may be a major influence for polar liquids such as liquid water and result in a modulation of the evaporative characteristics. It should also be noted that the thin film thickness profiles have been previously considered to be invariant beyond a critical channel radius of the order of ~ 2.5 µm[18].

Most importantly, the influence of the capillary pressure ($P_c$) in the liquid flow through the porous media, as well as in modifying the thin film curvature should also be considered. The $P_c$ is related to the $\sigma$ as well as the interfacial curvature ($K = y''(z)(1 + y'(z)^2)^{-\frac{3}{2}}$), through $P_c = \sigma K$, where $y''(z)$ and $y'(z)$ are the second and first derivatives of the thickness of the film with respect to the height ($z$) along the pore. Considering a very simple model of flow through a circular pipe in response to a capillary pressure difference ($P_c \sim \Delta P \sim \frac{4\sigma}{d}$) along the height of the pore channel, the flow rate (*m³/s*): $Q = \frac{\pi d^4}{128\mu} \frac{\Delta P}{H}$. For a $d$=253 µm, and $\Delta P$~ 1.15 kPa, the $Q$ was estimated to be ~ $10^{-8}$ m³/s or equivalently ~ $10^{-5}$ kg/s for water. From $Q$ as the product of the cross sectional area (=$A$) and the flow velocity (= $v_s$), we estimate a $v_s$ of ~ 0.2 m/s. While a greater $P_c$ (*e.g.,* due to a reduced *d*) would be expected to reduce the evaporative driving force due to increased liquid flow, it may also result in a flatter liquid profile in the evaporating thin film. Consequently, the thermal gradient could be reduced with the possibility of increased heat transfer.

The influence of the Capillary number $(Ca) = \mu v_s/\sigma$, which considers the relative effects of laminar flow (say, at a velocity $v_s$) *vs.* capillary flow is pertinent. Considering Poiseuille flow-like dynamics for water (with $\mu \sim 10^{-3}$ Pa.s and $\sigma \sim 72 \times 10^{-3}$ N/m), we estimate a *Ca* of the order of $10^{-3}$, for an ~250 μm pore diameter. Assuming heat dominated flow, where $v_s = h\Delta T/\rho_l L$, with *h* as the evaporative heat transfer coefficient of the heated water, and *ΔT* as the superheat[19] would yield an even smaller *Ca*, again indicating the dominance of capillary flow.

It is also relevant to note that due to the (i) variation with temperature along the pore, and (ii) relatively small depth (related to the thickness of the carbon foam) over which water flow occurs, that the surface tension driven convection (Marangoni effect) may also need to be considered. Marangoni convection could arise in the pores due to a temperature variation in the surface tension. For instance, a smaller difference of the temperatures, *e.g.,* between the thermocouple: TC1, and TC2 - see Fig. 1(a), was observed for the case of pure water (difference of ~ 2 °C) in contrast to the sample with 253 μm pore size (difference of ~ 15 °C). The pressure due to Marangoni flows, over a pore height of *H* (~ 1.2 cm) may be estimated through considering the temperature variation of the surface tension ($\frac{d\sigma}{dT} \sim \sim -2 \cdot 10^{-4} N/mK$, for water) as well as the thermal expansion coefficient[20] ($\alpha = \frac{1}{l}\frac{dl}{dT}$, $\sim 228 \cdot 10^{-6}/K$ at 22 °C, and $\sim 437 \cdot 10^{-6}/K$ at 47 °C for water, through:

$$\frac{d\sigma}{dl}\left(=\frac{d\sigma}{dT}\frac{dT}{dl}\frac{l}{l}\right) = \frac{d\sigma}{dT}\frac{1}{\alpha}\frac{1}{l} \qquad (1)$$

Consequently, a pressure variation from the bottom (at ~ 90 Pa) to the top of the sample (at ~ 40 Pa) drives the flows. However, the considered values of the Marangoni flow related pressure are at most 10% of the $P_c$ contributions.

Now considering the $\Delta m$ over a period of 1 hour of ~ 1.5 kg/m² for $d$=253 µm: Fig. 2(a), the equivalent evaporation rate over the carbon foam (covering the beaker of diameter ~ 3.6 cm) is of the order of 4.2·10$^{-7}$ kg/s – a factor of 25 lower than the value estimated through Poiseuille flow dynamics. Such a discrepancy may be ascribed to the tortuosity of the porous structure within the carbon foam. We note, for instance, that the ratio of the $P_c$ (equivalent to $\Delta P \sim \frac{4\sigma}{d}$) for the smallest average pore diameter of $d$=253 µm to the largest pore diameter of $d$=546 µm, in our study, is a factor of $\sqrt{3}$ higher compared to the respectively experimentally observed respective $\Delta m$ ratio (*i.e.,* 1.53 kg/m² for $d$ ~253 µm *vs.* 1.21 kg/m² for $d$ ~ 546 µm). As the diffusion length is proportional to the square root of the dimensionality, such a relation seems to indicate the importance of further considering the influence of pore ordering in evaporative processes. Moreover, more sophisticated mechanisms of heat transfer, *e.g.,* where vapor is emitted while water is simultaneously drawn into the internal cavities/tunnels[21] may be relevant but are difficult to apply in our present case due to the lack of precise knowledge of pore structure.

We then conclude that relatively minor contributions to evaporation arise from (a) the evaporative thin film region[16], *e.g.,* incorporating the triple-line region[22], and (b) thermocapillary convection due to Marangoni flows. Instead, capillary induced Poiseuille flows[23] seem to provide a rationale for the observed experimental results. Consequently, the efficiency of the water evaporation may be parameterized through the *Ca,* with a lower *Ca* being beneficial. Indeed, such a conclusion was previously drawn[24] in a study on the drying of porous media, through considering the formation and flow of macroscopic liquid films on the porous surface. However, even/uniform evaporation must be balanced with the lower throughput of the water vapor/steam generation.

The experimentally observed variation of the temperature along the height of the beaker indicates a progressive heating, which saturates as a function of time. Such behavior was analyzed as due to the competing effects of conduction and convection. From energy balance, a resultant evaporative heat flux from an effective surface area of $A_s$ due to the instantaneous temperature difference between the water in the carbon foam: at a temperature $T(t)$, and the ambient: at a temperature $T_a$, yields a convective heat flux ($= hA_s[T(t)-T_a]$). Such a flux is assumed to be due to the water mass loss ($=d(\Delta m)/dt = \rho VC\, dT(t)/dt$), where $\rho$ is the water density, $V$ is the volume, and $C$ the specific heat of the water in the porous medium. Consequently, it may be derived, for an initial water temperature: $T_i$, that:

$$\frac{T(t)-T_a}{T_i-T_a} = exp\left(-\frac{hA_s}{\rho VC}t\right) \qquad (2)$$

Plots of the TC1 temperature: $T(t)$ fit to the above relation are shown in Fig. 3(a), comparing a bare water surface and the studied porous surfaces (see supplementary material[29] in Section S2). We noted that the plots in Fig. 3(a), could be distinguished through the increased $hA_s/\rho VC$ ratio as a function of the pore size. For instance, in the case of pure water, we obtained a value for the ratio of ~ 9.3 x $10^{-4}$, yielding a value of $h$ ~ 47 W/m$^2$K, while for the smallest $d$ ~ 253 $\mu$m, the corresponding ratio is ~ 8.3 x $10^{-3}$, yielding a value of $h$ enhanced almost by an *order of magnitude* to ~ 420 W/m$^2$K. The inverse of the ratio is commonly considered in terms of a thermal time constant (*i.e.,* $\tau = \rho VC/hA_s$), as treated through a lumped element approach[20], with $1/hA_s$ indicating a resistance to convective heat transfer and $\rho VC$ being the capacitance of the water in the porous medium, and is plotted in Fig. 3(b). Given that the porosity of all the studied carbon foams is ~ 97%, we may interpret the varying time constants as due to the increase of the $h$ with reduced pore size. Indeed, the possibility of rough enhanced surfaces for increased heat transfer has been considered in much detail[25,26].

In addition to the influence of pore size on regulating the rate of water evaporation through convective methodologies, we attempted to investigate the role of the materials surface. The related specific characteristics have been focused on the $P_d$ but may play a larger role. For example, it may be thought that increased hydro-/phobicity (/- philicity) would retract (/advance) the *evaporative thin film* region over which evaporation effectively occurs. Consequently, it was hypothesized that inducing hydrophilic character onto the carbon foam surface would effectively flatten the film profile[18], over the meniscus region, due to creep of the water up the sides of the pores, and enhance the $P_c$. To this end, the carbon foams used for the porous media were subject to nitric acid ($HNO_3$) treatment for two hours, under sonication, which has been shown previously to oxidize the graphitic surface and introduce hydrophilic functional groups[27]. An increasing molarity of the acid was used to enhance hydrophilic character. It was then observed that the $\Delta m$ was increased, with a reduced waiting time observed at any given pore size, with increased molarity: Fig. 4(a) indicates the data for $d$ = 253 µm; – also see supplementary material[29] in Section S3). An increased effective hydrophilic interaction seems to be implied, contributing to enhanced evaporation akin to effects related to a reduced pore size, *cf.,* Fig. 2(a). We depict the inferred variation in the meniscus shape indicating an *increased* evaporative thin film region at increased surface hydrophilicity of the carbon foam surface in Fig. 4(b), which in turn leads to enhanced evaporative mass flux.

In summary, we have shown that enhanced capillary pressures, either due to a decreased pore size or the chemical modification of the surfaces would serve to increase the evaporative efficiency of water through light absorbing (see supplementary material[29] in Section S4, where it is indicated that the absorption could be of the order of 93%-95%) porous media. The proposed media may also be used for desalination (see supplementary material[29] in Section S5) where the larger $\sigma$ may be of advantage; additionally, a decreased interfacial thermal resistance[28] in ionic solutions may also be beneficial.

We would like to thank Oscar Rios at UC, San Diego for help in the 3D printing of the aerogel insulation container. The authors acknowledge financial support from the Defense Advanced Research Projects Agency (DARPA: W911NF-15-2-0122) and the National Science Foundation (NSF: CMMI 1246800).

**References**

[1] K. Sampathkumar, T.V. Arjunan, P. Pitchandi, and P. Senthilkumar, Renew. Sustain. Energy Rev. **14**, 1503 (2010).

[2] M. Thirugnanasambandam, S. Iniyan, and R. Goic, Renew. Sustain. Energy Rev. **14**, 312 (2010).

[3] N.S. Lewis, Science (80-. ). **351**, aad1920 (2016).

[4] H. Ghasemi, G. Ni, A.M. Marconnet, J. Loomis, S. Yerci, N. Miljkovic, and G. Chen, Nat. Commun. **5**, 4449 (2014).

[5] L. Zhang, B. Tang, J. Wu, R. Li, and P. Wang, Adv. Mater. **27**, 4889 (2015).

[6] K. Bae, G. Kang, S.K. Cho, W. Park, K. Kim, and W.J. Padilla, Nat. Commun. **6**, 10103 (2015).

[7] L. Zhou, Y. Tan, D. Ji, B. Zhu, P. Zhang, J. Xu, Q. Gan, Z. Yu, and J. Zhu, Sci. Adv. **2**, e1501227 (2016).

[8] S. Ishii, R.P. Sugavaneshwar, and T. Nagao, J. Phys. Chem. C **120**, 2343 (2016).

[9] S. Ishii, R.P. Sugavaneshwar, K. Chen, T.D. Dao, and T. Nagao, Opt. Mater. Express **6**, 640 (2016).

[10] X. Wang, G. Ou, N. Wang, and H. Wu, ACS Appl. Mater. Interfaces **8**, 9194 (2016).

[11] M.S. Zielinski, J.-W. Choi, T. La Grange, M. Modestino, S.M.H. Hashemi, Y. Pu, S. Birkhold, J.A. Hubbell, and D. Psaltis, Nano Lett. **16**, 2159 (2016).

[12] O. Neumann, A.D. Neumann, E. Silva, C. Ayala-Orozco, S. Tian, P. Nordlander, and N.J. Halas, Nano Lett. **15**, 7880 (2015).

[13] S. Lee, R.A. Taylor, L. Dai, R. Prasher, and P.E. Phelan, Mater. Res. Express **2**, 065004 (2015).

[14] S.J.S. Morris, J. Fluid Mech. **494**, 297 (2003).

[15] S. Yu, Y. Zhang, H. Duan, Y. Liu, X. Quan, P. Tao, W. Shang, J. Wu, C. Song, and T. Deng, Sci. Rep. **5**, 13600 (2015).

[16] M. Potash and P.. Wayner, Int. J. Heat Mass Transf. **15**, 1851 (1972).

[17] J.N. Israelachvili, *Intermolecular and Surface Forces,* 3rd ed. (Academic Press, San Diego, 2011).

[18] H. Wang, S. V. Garimella, and J.Y. Murthy, Int. J. Heat Mass Transf. **50**, 3933 (2007).

[19] H.K. Cammenga, in *Curr. Top. Mater. Sci.* (North Holland Publishing Company, New York, NY, 1980), pp. 335–446.

[20] T.L. Bergman, A.S. Lavine, F.P. Incropera, and D.P. Dewitt, *Introduction to Heat Transfer* (John Wiley Inc., Hoboken, NJ, 2011).

[21] W. Nakayama, T. Daikoku, H. Kuwahara, and T. Nakajima, J. Heat Transfer **102**, 451 (1980).

[22] H.K. Dhavaleswarapu, J.Y. Murthy, and S. V. Garimella, Int. J. Heat Mass Transf. **55**, 915 (2012).

[23] P.K. Kundu and I.M. Cohen, *Fluid Mechanics*, 4th ed. (Academic Press, San Diego, CA, 2008).

[24] A.G. Yiotis, A.G. Boudouvis, A.K. Stubos, I.N. Tsimpanogiannis, and Y.C. Yortsos, Phys. Rev. E. Stat. Nonlin. Soft Matter Phys. **68**, 037303 (2003).

[25] A.E. Bergles, J. Heat Transfer **110**, 1082 (1988).

[26] L. Burmeister, *Convective Heat Transfer*, 2nd ed. (John Wiley & Sons Inc., New York, NY, 1993).

[27] E. Bouleghlimat, P.R. Davies, R.J. Davies, R. Howarth, J. Kulhavy, and D.J. Morgan, Carbon N. Y. **61**, 124 (2013).

[28] S. Baral, A.J. Green, and H.H. Richardson, MRS Proc. **1779**, 33 (2015).

[29] See supplementary material at [URL will be inserted by AIP] for (i) Pore size distribution for the carbon foam samples, (ii) Temperature profiles for the carbon foam samples, (iii) Influence of nitric acid treatment on modulating the hydrophilic character of carbon foams, (iv) Optical characteristics of reflection and transmission for the prepared carbon foams, (v) Desalination characteristics of the prepared carbon foams

Figure Captions

# Figure 1

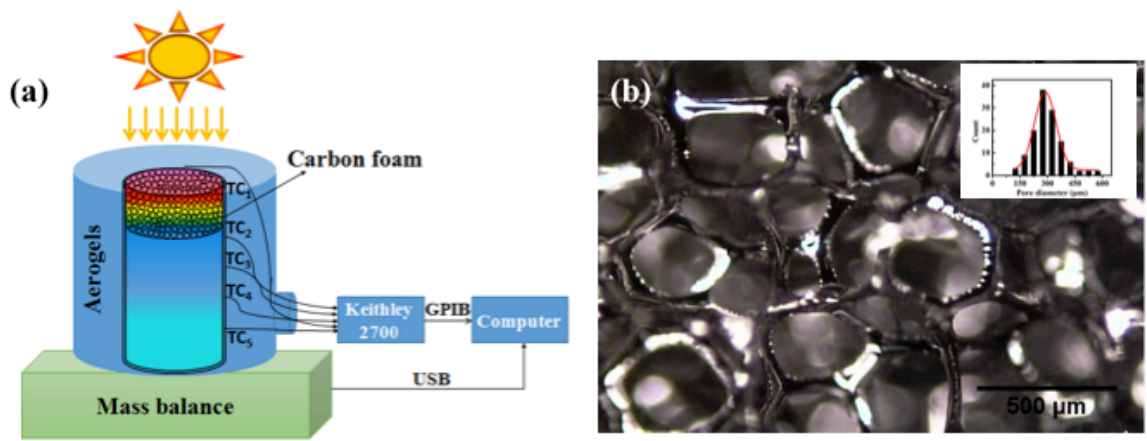

Canbazoglu, *et al*

---

**Figure 1 (a)** Experimental setup, related to solar radiation mediated evaporation of water. Asolar light simulator was used as an input power source for heating the water in the porous carbon foams, insulated laterally through a low thermal conductivity, aerogel infiltrated, chamber. The thermocouples (TC) probe the water and vapor temperature along the height of the beaker, while the mass balance at the bottom was calibrated for the mass loss ($\Delta m$) due to water evaporation. **(b)** A SEM (scanning electron microscope) micrograph of the top surface of a typical porous structure of the carbon foams used in the experiments. The *inset* indicates a pore size variation distribution for a 80 PPI sample (with a mean pore size of 288 μm and a standard deviation of ~ 59 μm).

## Figure 2

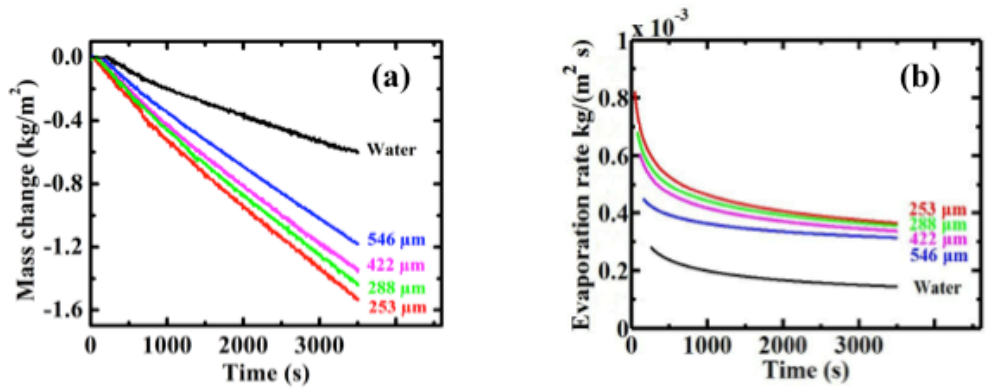

Canbazoglu, *et al*

**Figure 2** **(a)** The variation of the area averaged evaporative mass flux loss ($\Delta m$) with time ($t$) indicates the greater effectiveness of smaller pore sizes, compared to bare water. **(b)** The derived rate of the mass loss ($=d(\Delta m)/dt$), indicates non-uniform evaporation and that a steady state is reached at longer times, with a larger time at smaller $d$.

## Figure 3

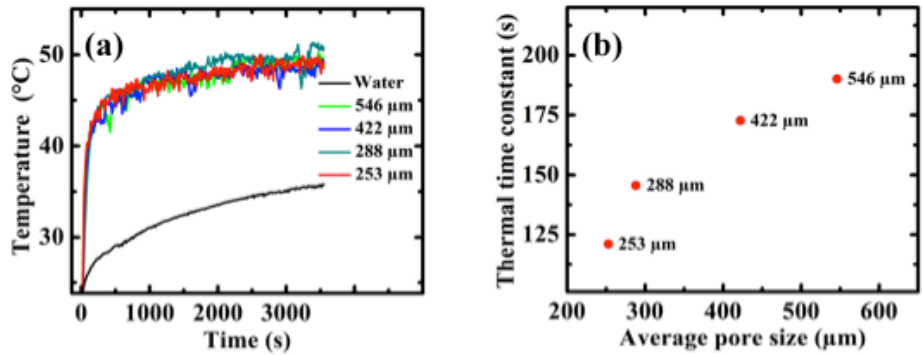

Canbazoglu, *et al*

**Figure 3 (a)** The observed variation of the TC1 temperature from a bare water surface compared to that in the studied porous surfaces, indicates competing effects of conductive and convective heat transfer, **(b)** A plot of the deduced thermal time constant, for water evaporation, as a function of the pore size.

**Figure 4**

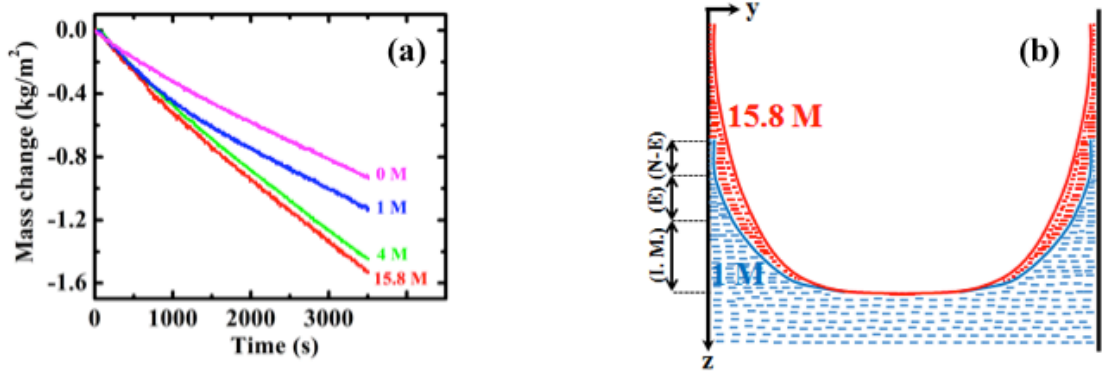

Canbazoglu, *et al*

**Figure 4 (a)** The variation of the evaporative mass flux loss ($\Delta m$) with time as a function of increased molarity of nitric acid, which was used to enhance the hydrophilic character of the carbon foam surface. **(b)** The inferred variation in the meniscus shape indicating an *increased* evaporative thin film region, due to enhanced surface hydrophilicity through larger nitric acid molarity (15.8 M *vs.* 1 M) treatment of the carbon foam surface. The **I.M**. (*Intrinsic meniscus*), **E**. (*evaporating thin film*) and **N-E** (*non-evaporating region*) regions are shown.